# Insider Detection using Deep Autoencoder and Variational Autoencoder Neural Networks

Efthimios Pantelidis*, Gueltoum Bendiab†, Stavros Shiaeles†, Nicholas Kolokotronis‡
* Faculty of Science and Applied Sciences, Open University of Cyprus (OUC)
Nicosia, Cyprus, efthymios.pantelides@st.ouc.ac.cy
†Cyber Security Research Group, University of Portsmouth, PO1 2UP, Portsmouth, UK
gueltoum.bendiab@port.ac.uk, sshiaeles@ieee.org
‡Department of Informatics and Telecommunications, University of Peloponnese
22131 Tripolis, Greece, nkolok@uop.gr

*Abstract*—Insider attacks are one of the most challenging cybersecurity issues for companies, businesses and critical infrastructures. Despite the implemented perimeter defences, the risk of this kind of attacks still very high. In fact, the detection of insider attacks is a very complicated security task and presents a serious challenge to the research community. In this paper, we aim to address this issue by using deep learning algorithms Autoencoder and Variational Autoencoder deep. We will especially investigate the usefulness of applying these algorithms to automatically defend against potential internal threats, without human intervention. The effectiveness of these two models is evaluated on the public dataset CERT dataset (CERT r4.2). This version of the CERT Insider Threat Test dataset includes both benign and malicious activities generated from 1000 simulated users. The comparison results with other models show that the Variational Autoencoder neural network provides the best overall performance with a greater detection accuracy and a reasonable false positive rate.

## I. INTRODUCTION

Insider threat is currently one of the most significant security problems for institutions, businesses, and government agencies [1], [2]. This threat usually arises from users with authorised access to an organisation' system who use that access either maliciously or unintentionally, to cause data breaches and harms to the organisation [3], [4]. This user could be a current employee in the organisation or other business associate such as a consultant, a broad member, a former employee, etc. The insider actor can perform internal attacks intentionally for malicious purposes like espionage, disclosure of sensitive information or intellectual property theft. This kind of internal attacks usually implicates a current or former employee or any other business partners that have privileged accounts within the network of an organisation, and who misuses this access [4]. Insider threat could also be performed by exploiting a negligent insider who does not follow proper security rules, such as an employee who did not change a default password or fall victim to a phishing attempt [5]. In either case, negligence is often considered the most expensive type of insider risk [6]. According to Verizon Data Breach Investigations Report, more than 30% of data breaches in 2020 involve internal actors [7], and more than 40% of these violations, usually take months or even years to be identified. Another recent report by the Pomeron institute announced that the average annual cost of insider threat was $11.45 million in 2020, with an increase of 31% compared to 2019 [8].

Despite the implemented security measures, the risk of insider attacks remains very high. In fact, these attacks are especially more dangerous than outside attacks because an insider already has direct access to the organisation system, its network and sensitive data [2], [3]. In addition, it is very difficult to detect insider threats because they can easily evade existing security mechanisms. Moreover, it is very difficult to distinguish between a legitimate user's activity and potentially malicious activity, especially for those with privileged access [9]. While this issue has been explored for a long time in both industrial and research communities, most proposed methods rely on feature engineering, which is difficult and time-consuming. In addition, most of them cannot correctly capture the behaviour difference between malicious and normal user activities due to numerous difficulties related to the related data. Deep learning techniques give a new powerful paradigm that can automatically discover the features needed for insider threat detection. Although some progress has been done in this area, the topic of applying deep learning for insider threat detection is not deeply investigated [6], [10]. Therefore, this paper studies the application of the deep neural networks Autoencoder (AE) and Variational Autoencoder (VAE) in detecting malicious insiders automatically, without human intervention. These two deep neural networks have proved their effectiveness in anomaly detection and provided promising results in various research work [11], [12]. According to [6], AE and VAE can learn different levels of representations from the input data based on the multi-layer structures. Moreover, The VAE could generate new data from the source dataset [6]. In this study, we have used the Python programming language with the Keras library and the TensorFlow environment to implement the AE and VAE neural networks. While the validation of the implemented models is performed on the public CERT (version r4.2). This version of the CERT- dataset includes both normal and malicious activities that are generated from 1000 simulated employees. The comparison results with our previous models in [11], [12] indicate that the Variational Autoencoder neural network provides the best overall accuracy in detecting internal threats with a lower false-positive rate. The rest of this paper is structured as follow: Section 2 reviews the related work in the area of insider threat. Section 3 explains the proposed method to detect insider threats and specific parameters. Section 4 presents the testing design and results, and Section 5 presumes the paper along with some future directions.

## II. RELATED WORK

With the increasing number of incidents in recent years, several research studies have attempted to design new techniques to solve the problem of insider threat detection [6], [13] and several techniques have been proposed in the literature. A survey in [2] has studied the most recent solutions in this field and proposed a novel categorisation of insider threat- related work based on the techniques used in the detection. The authors concluded that machine learning-based detection techniques are the most powerful for solving the problem of insider threat. In this technique, a user normal profile is built based on their normal behaviour. In this case, the anomalies are identified as deviations from the normal behaviour [2], [5], [14]. In this context, a variety of machine learning algorithms have been used for insider threat detection. For instance, authors in [14] introduced an automated system that has the ability to detect insider threat based on the user's profile. Each user's profile is built based on a three-structure approach that involves the details of their activities and job role. The created profile is then used for comparison with other users' activities of the same role. Significant deviations in the user's activities will be considered as an anomaly and potential internal threat. In a recent work [15], authors proposed a new trust management model to improve the protection of collaborative Intrusion Detection Systems (IDSs) against insider threats. The proposed model used a supervised machine learning technique to automatically assign a trust value, known as intrusion sensitivity, to each IDS based on expert knowledge. Then, the trustworthiness of each IDS in the system is evaluated using this trust value. IDS with a low trust level is considered a potential insider threat actor. In this work, the authors investigated the performance of three machine learning algorithms, k-nearest Neighbours algorithm (KNN), back-propagation neural networks (BPNN) and decision tree (DT), in assigning trust value under various attack scenarios in a real WSN (Wireless Sensor Network). In another work [16], authors examined the efficiency of using supervised machine learning and data mining to identify insider threats. The proposed models have been evaluated on a dataset that includes activities of 24 users generated over five days of spam. The machine learning algorithms used in the experiments are Adaboost, Naive Bayes (NB), Logistic Regression (LR), KNN, Linear Regression (LR) and Support Vector Machine (SVM). The authors found that Adaboost has exceeded other algorithms with 98.3% accuracy in identifying malicious emails. In [17], the authors used a Hidden Markov Model (HMM) to build a user profile by capturing his normal activity per week. Then, this profile is used to detect significant deviations from the normal behaviour. In the same context, in [18], authors tested the efficiency of different machine learning algorithms in detecting anomalies and early quitter indicators, where both indicate a potential internal attack. The proposed detection models are tested on a dataset that includes activities of 5,500 simulated users.

Many other recent detection approaches have been used deep learning algorithms. The main advantage of these techniques is that they can automatically discover the features needed for the detection of insider attacks from big data, with- out being particularly programmed [19], [20]. For instance, the Recurrent Neural Network (RNN) has been used in many research works to model the user activities for insider threat detection [19], [21]. Especially, Long Short-Term Memory (LSTM) [21] and Gated Recurrent Unit (GRU) are the two main variants of the RRN that have been widely employed to model the user activities and building a user profile [10], [19], [22], [23]. For instance, in [10], authors employed Long Short-Term Memory (LSTM) to extract temporal features for building the user profile and the Convolutional Neural Network (CNN) to detect insider threat using the extracted features. In [23], authors proposed an online anomaly detection approach using RRN to produce anomaly scores. Convolutional Neu- ral Network (CNN), which gained great success in images classification [24], has also been used to detect insider threat by analysing images generated from the user mouse bio- behavioural characteristics [24]. The proposed approach can perform continuous identity authentication of computer' users with a false acceptance rate of 2.94% and a false rejection rate of 2.28%. In [25] Deep Autoencoders (AE) have been used to detect insider threat. In this work, each AE is trained using a certain category of audit data, which indicates the user's normal behaviour. The authors claimed that the proposed detection system is able to detect all malicious insider actions. However, this model has a high false-positive rate. In a similar work [26], authors implemented a Deep Neural Network (DNN) model based on the auto-encoder NN to detect insider threat. The validation of this system is performed with a real- world data set that includes 3.6 billion log files and 70.2 million entities. This system achieved good results; however, the modelling of the user activities overtime was not clear.

## III. PROPOSED METHODOLOGY

As mentioned in the previous section, the main goal of this work is to evaluate the effectiveness of the deep learning algorithms AE and VAE in detecting insider attacks, without human intervention. The main idea of the paper is presented in Fig. 1. As shown in the figure, the proposed system involves two main steps in order to detect user activities that refer to a potential insider attack. The first step involves data collection from the public dataset CERT r4.2 [11]. This dataset was created by SEI (Software Engineering Institute), in 2010, for insider threat tests purposes and it is publicly available. It comprises log files that record an activity including eighteen

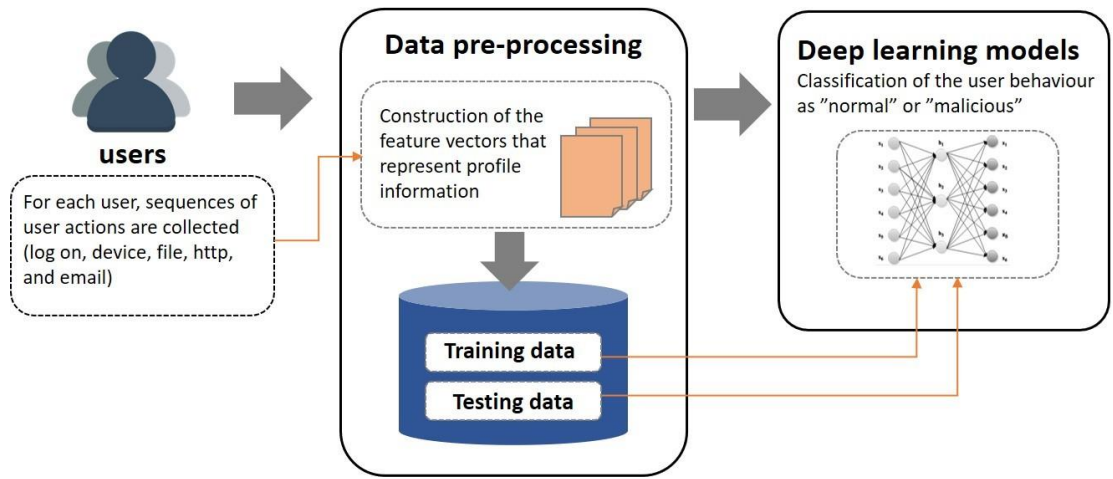

Fig. 1. Overview of the proposed methodology

(18) months, collected between 01.01.2010 and 31.05.2011 [11]. For each user in the dataset, the inputs involve login traces, files used by the users, emails sent and received, web browsing records, removable media used, and employee role within the organisation. Then, in the next step of data collection and processing, the collected data are processed to create feature vectors representing profile information and converted into a format (i.e., numerical format) that can be fed into the learning algorithms. Created feature vectors are then fed to the implemented neural networks autoencoder (AE) and variational autoencoder (VAE) to perform predictions and classify user activities as normal or anomaly. The following sections provide detailed information about the steps we followed to complete the process and classify user behaviour as "normal" or "malicious".

*A. Dataset and Data Collection*

As mentioned before, data collection is the first step in the proposed system for insider threat detection. This step is crucial because proper data collection and processing allows a successful application of the neural networks models AE and VAE with a higher accuracy rate and lower false alarms. There are actually serval public datasets that have been created for insider threat tests purposes. However, in this paper, we have used the public insider threat dataset created by SEI (Software Engineering Institute) [21]. The institute proposed a collection of ten different test datasets (r1, r2, r3.1, r3.2, r4.1, r4.2, r5.1, r5.2, r6.1, r 6.2) that provide both background and malicious actor synthetic data. In this work, we have chosen version r4.2 of the CERT dataset [21]. This version of the dataset comprises many samples that are collected from three different scenarios of insider threat. The dataset is "free of privacy and restriction limitations". It includes both benign and malicious user activities generated from 1000 simulated employees. The CERT r4.2 dataset is divided into seven log files, which record users activity covering eighteen (18) months, collected between 01.01.2010 and 31.05.2011. The five log files are presented in Table I.

TABLE I
CSV LOG FILES OF THE CERT R4.2 DATASET

| File Name | Description |
|---|---|
| logon.csv | Records the user's input and output |
| device.csv | Records USB connection and disconnection |
| http.csv | Records internet usage |
| email.csv | Records e-mail usage |
| file.csv | Records saving files to removable devices |
| psychometric.csv | Psychometric results |
| ldap.csv | Users of the organisation and their roles |

The version r4.2 of the CERT dataset contains three main insider threat scenarios, where the insider actor is a current employee in an organisation. In the first scenario, the employee has logged into the organisation's system after working hours in order to collect and store sensitive data from the organization' system on a removable disk. Then, he uploaded the collected data to the "wikileaks.org" website. After that, the employee left the organization. In the second scenario, the employee starts to look for a new job by browsing job websites and requesting employment from an adversary to the organisation. Before moving to the adversary company, he used a USB thrum drive to steal sensitive data from the organisation system. Finally, in the third scenario, the current employee is a system administrator in the company with privileged access rights. The employee becomes disappointed for some reasons. Consequently, he downloaded a keylogger and used a USB stick to transfer it to his supervisor's machine. Then, he logged in as a supervisor and sent an alarming mass email, which caused panic in the organisation. The system administrator left the organisation shortly. Around 20 GB of employee activity logs have been generated from these three scenarios that involved 30 malicious employees.

## B. Data Pre-Processing

In the data processing step, the raw input files are imported from the CERT r4.2 and processed. Initially, due to the large volume of data and high memory requirements, we have created sample files of 5000 entries per column for each of the CSV files, logon, device, file, HTTP, and email. During this process, we have extracted seven features that represent the possible user activities in the system. These features are Logon/Logoff activities, connect/disconnect thumb drive, sent/received e-mails, file processing and Internet browsing activities. All these features (activities) are collected in a single CSV file with entries from 1000 rows. The characteristics used to create this file are "day", "time", "Logon", "Logoff", "Connect", "Disconnect", "email", "file", and "HTTP". Because the deep learning models require all input and output variables to be numeric, the non-numerical features have been converted to numeric values (Integer) as follows: The seven users' activities (logon, logoff, connect, disconnect, email, file and HTTP) are mapped to integer numbers from 1 to 7. The "date" column has been converted to "day" and "time". After that, the hot encoding method is used to convert all the inputs to binary 0 and 1. Table II presents the numerical values of the extracted features at the pre-processing stage.

TABLE II
SUMMARY OF THE EXTRACTED FEATURES AT PRE-PROCESSING PHASE.

| Feature | Possible Values |
|---------|-----------------|
| Day | 0, 1, 2, 3, 4, 5, 6 |
| Time | 1, 2, 3, 4, . . ., 24 |
| Activities | "Logon" =1, "Logoff" =2, "Connect" =3, "Disconnect" =4, "email" =5, "file" =6, "http" =7 |

Finally, we have divided the obtained dataset into two subsets, which are training and testing sets. The training dataset, which includes 75% of all data, is used for training the AE and VAE models with labelled data. For that, a set of usernames of the malicious insiders is created according to the scenarios in the CERT r4.2 dataset. Then, a new column has been added to the CSV file, called "insider" with binary value. The value of this feature is set to 1 if the user is an insider. Otherwise, it is set to 0. While the testing dataset, which contains 25% of all data, is used to validate the efficacy of the implemented models in predicting the testing data.

## C. Implementing AE and VAE algorithms to detect anomalies

During the testing stage, we have built the structure of the AE and VAE by using used the Python programming language with the Keras library and the TensorFlow environment. After that, we have identified the best values of the related parameters including the number of layers for each model and for each layer, the best activation function to use. The structure of the Autoencoder neural network is constituted of two main connected networks, an encoder and a decoder network. The encoder network is used to learn how to read the input data and compress it to an internal representation defined by the bottleneck layer. The output of this network is then used by the decoder network to reconstruct the initial input data [27]. After sufficient training of the encoder network, the decoder can be dropped. Finally, we keep only the trained encoder network, which will be used to compress samples of input data to output vectors. In the implementation step of this neural network, we have built an input level (i.e., 50), encoder network, decoder network and output level. The encoder network involves two hidden layers. The first layer with two times the number of inputs (e.g., 100) and the second with the same number of inputs (i.e., 50), succeeded by the bottleneck layer with the same number of inputs as the dataset (i.e., 50). The decoder network was implemented with a similar architecture, although in reverse. In addition, we have determined the activation function at the levels (tanh, relu) and the dense layer.

The Variational Autoencoders (VAEs) are generative models which are able to generate new output data that are highly similar to the training data 2. Just like AE, the architecture of the VAE has also two main networks; an encoder and a decoder. The encoder network will be trained to minimise the reconstruction error between the encoded-decoded data and the initial data. However, its encoder network outputs two different vectors of size n. A vector of means called $\mu$, and another vector of standard deviations called $\sigma$ [12]. For the implementation of the VAE model, we have created the decoder and the encoder networks in addition to the input and output layers. While the optimization is performed with the Nesterov Adam optimizer (NADAM).

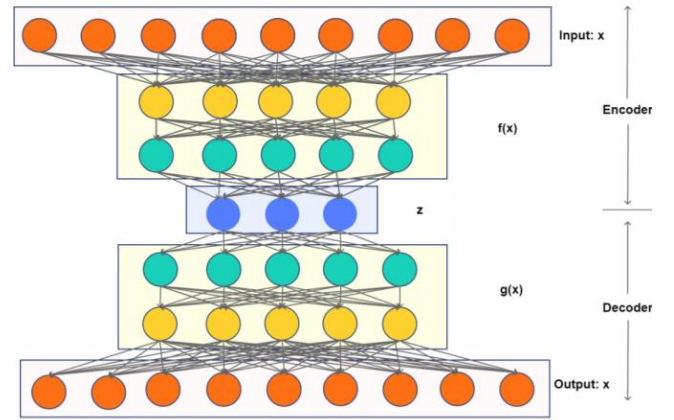

Fig. 2. Variational Autoencoder structure [28]

## IV. SYSTEM IMPLEMENTATION & TESTING

The experiments were carried out in a virtualised environment. The AE and VEA neural networks were implemented using the Python programming language, TensorFlow library and the open-source library Keras that provides a Python interface for artificial neural networks and the TensorFlow library. The metrics used to evaluate the effectiveness of each model are:

Accuracy (A): refers to the percentage of all correctly classified activities for the test data. It can be computed using the following equation:

$$Accuracy(A) = \frac{TP + TN}{TP + TN + FP + FN} \quad (1)$$

In these experiments, malicious activities represent positive instances, while normal activities represent negative instance. True Positive (TP) is the number of positive instances that have been correctly classified. False Positive (FP) is the number of abnormal instances that have been incorrectly classified. True Negative (TN) is the number of negative instances that have been correctly classified. False Negative (FN) is the number of normal activities that have been incorrectly classified.

Precision (P) refers to the percentage of positive instances that have been correctly classified.

$$Precision\ (P) = \frac{TP}{TP + FP} \quad (2)$$

Recall (R) provides the number of negative instances that were correctly classified. It indicates the efficiency of the prediction model in detecting relevant data.

$$Recall\ (R) = \frac{TP}{TP + FN} \quad (3)$$

F1-score is a weighted percentage between precision and recall.

$$F1 - score\ (RF1) = 100 \times \frac{P \times R}{P + R} \quad (3)$$

### A. Results and discussion

Before running the tests, the training of the AE neural network is done for 30 epochs with a batch size of 256. For the Variational Autoencoder (VAE) neural network, the training is done for 1000 epochs with a batch size of 128, while the dimension of the latent space is set to 2. The optimal values of these parameters have been identified by using the learning rate function (LRFinder) [29]. After that, many tests were carried out on the trained models to determine their performance. As shown in Fig. 3, the VAE clearly outperforms the AE model, where it achieved higher detection performance with an overall accuracy rate of 96%, precision (92%) and recall (96%). Based on these results, the F1-score value is 94%. These values illustrate the efficiency of the VAE in the correct classification of most of the samples. On the other hand, the EA allows slightly lower accuracy detection (95%) and recall (95%) at a cost of a higher false alarm rate (precision rate is 90%).

### B. Comparison with other models

The implemented models are compared to our previous work in [11], [12], based on the predefined metrics. In [11], we used the Convolutional Neural Network (CNN) algorithm to identify potential insider threats by using a visual representation of the activity report. The training and testing of the CNN algorithm are done with a dataset of 860 2D images.

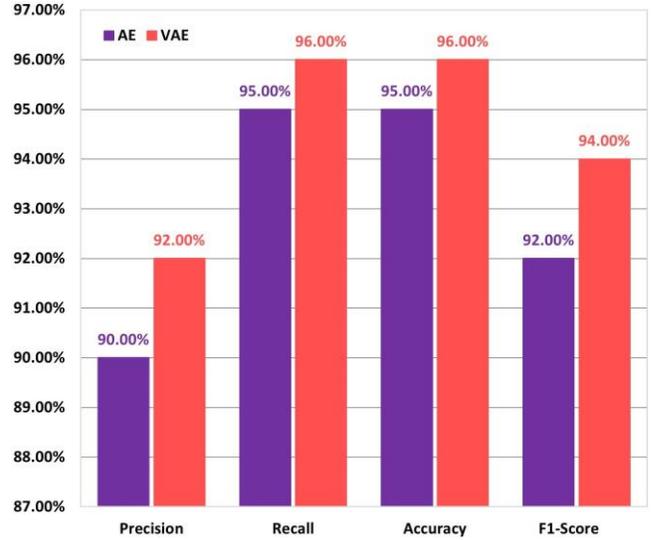

Fig. 3. Evaluation metrics results

More specifically, 80% of the samples (769 images) were labelled as containing malicious activity and 430 images were labelled as normal. In [12], we examined the efficiency of swarm intelligence algorithms in feature selection optimization. This will greatly enhance the performance of the machine learning model Local Outlier Factor (LOF) in detecting inside threats. In this study, we compare the Bio-inspired model with the nine features "day", "time", "Logon", "Logoff", "Connect", "Disconnect", "email", "file", and "HTTP". The same dataset (CERT dataset) was used for training and testing all the algorithms. The evaluation results are reported in Table III.

TABLE III
COMPARISON WITH OTHER LEARNING ALGORITHMS

|  | Precision | Recall | Accuracy | F1-Score |
|---|---|---|---|---|
| Autoencoder (AE) | 90% | 95% | 95% | 92% |
| Variational Autoencoder (VAE) | 92% | 96% | 96% | 94% |
| CNN [11] | n/a | n/a | 90% | n/a |
| Bio-Inspired models [12] (9 features) | n/a | 100% | n/a | 70% |

As shown in Table III, the VAE neural network has achieved the best overall performance compared to other algorithms, with higher accuracy (96%) and recall (96%), and lower false alarms. The CNN model [11] has also achieved very high sensitivity (recall), with a percentage that reached 100%. However, the lack of values for all the evaluation metrics [11], [12] does not help to have a clear view of their performances. Thus, the two models in [11], [12] need to be studied further in future work.

## V. CONCLUSION

Insider threat is one of the most complicated security issues that cause significant loss to organisations and businesses. In this work, we have examined the utility of using deep learning techniques to detect insider threats, without human intervention. For that, we have implemented two deep neural networks autoencoder and Variational Autoencoder to check their efficiency in detecting insider threats automatically. The tests were performed on the public CERT dataset (CERT r4.2). From the tests' results and comparison with other deep learning algorithms, the Variational Autoencoder neural network has proved that it is the most effective in identifying internal threats with an overall accuracy of 96.00%. On the other hand, Autoencoder allows slightly better insider threat detection performance (accuracy 95%) at a cost of higher false alarm rates.

In the future, we intend to perform more experiments on the implemented models by using more data to accurately train and test the neural networks. This can greatly improve the overall performance of the proposed models. We also intend to compare the results of our models with the most relevant work in this field and to consider different types of metrics to evaluate them. In this context, it would be good to use the CERT dataset with other deep learning algorithms and compare the results. We are also planning to apply these neural networks to other problems related to cybersecurity, therefore not only to the identification of internal malicious attacks but also to external ones.

## ACKNOWLEDGMENT

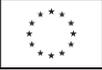
This project has received funding from the European Union's Horizon 2020 research and innovation programme under grant agreement no. 786698. The work reflects only the authors' view and the Agency is not responsible for any use that may be made of the information it contains.